\documentclass[aps,prl,superscriptaddress,twocolumn,nofootinbib,floatfix,notitlepage]{revtex4-1}

\usepackage{graphicx}
\usepackage{amsmath,amsfonts,amssymb}
\usepackage{hyperref}
\usepackage{color}
\usepackage{verbatim}

\newcommand{\be}{\begin{equation}}
\newcommand{\ee}{\end{equation}}
\newcommand{\bea}{\begin{equation}\begin{aligned}}
\newcommand{\eea}{\end{aligned}\end{equation}}
\newcommand{\td}{{\rm d}}

\begin{document}

\title{Impact of LIGO-Virgo black hole binaries on gravitational wave background searches}

\author{Marek Lewicki}
\email{marek.lewicki@fuw.edu.pl}
\affiliation{Faculty of Physics, University of Warsaw ul.\ Pasteura 5, 02-093 Warsaw, Poland}

\author{Ville Vaskonen}
\email{vvaskonen@ifae.es}
\affiliation{Institut de Fisica d'Altes Energies (IFAE), The Barcelona Institute of Science and Technology, Campus UAB, 08193 Bellaterra (Barcelona), Spain}

\begin{abstract}
We study the impact of the black hole binary population currently probed by LIGO-Virgo on future searches for the primordial gravitational wave background. We estimate the foreground generated by the binaries using the observed event rate and a simple modeling of the black hole population. We subtract individually resolvable binaries from the foreground and utilize Fisher analysis to derive sensitivity curves for power-law signals including these astrophysical foregrounds. Even with optimistic assumptions, we find that the reach of future experiments will be severely reduced.
\end{abstract}

\maketitle

\begin{figure}
    \includegraphics[width=0.48\textwidth]{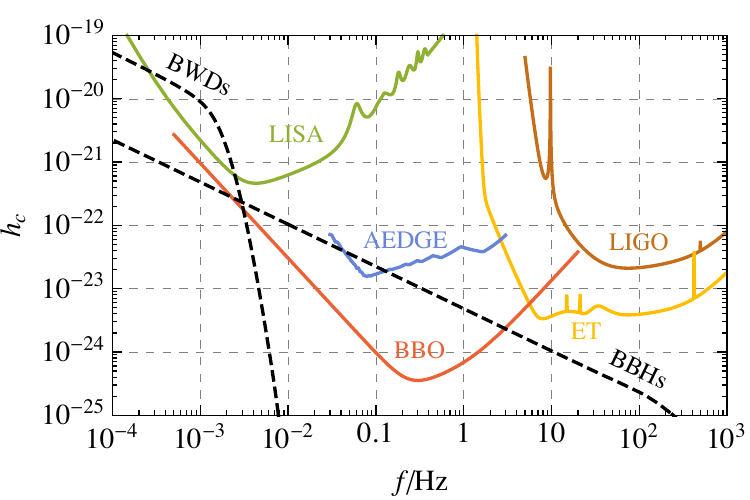}
    \caption{The instantaneous GW strain sensitivities of LIGO~\cite{LIGOScientific:2014pky}, ET~\cite{Hild:2010id}, LISA~\cite{LISA:2017pwj,Robson:2018ifk}, AEDGE~\cite{AEDGE:2019nxb,Canuel:2019abg,Badurina:2021rgt} and BBO~\cite{Crowder:2005nr}. The black dashed curves show the GW foregrounds from binary white dwarfs (BWDs)~\cite{Cornish:2017vip,Robson:2018ifk} and from the binary black hole (BBH) population currently probed by LIGO-Virgo~\cite{LIGOScientific:2020kqk,LIGOScientific:2021psn}.}
    \label{fig:sens}
\end{figure}

\noindent\textbf{Introduction} -- Various cosmological processes may have generated gravitational waves (GWs) that contribute to the stochastic GW background~\cite{Caprini:2015zlo,Bartolo:2016ami,Caprini:2018mtu,Auclair:2019wcv,Caprini:2019egz}. If sufficiently strong, this background can be probed with GW detectors. The non-observation of such background with LIGO-Virgo detectors~\cite{KAGRA:2021kbb} translates into mild constraints on the possible cosmological GW sources~\cite{LIGOScientific:2021nrg,Romero:2021kby,Romero-Rodriguez:2021aws}. As shown in Fig.~\ref{fig:sens}, many new experimental programs, built on the success of currently running LIGO-Virgo network, are expected to enter GW search in the next decades. With tremendous sensitivity improvements, these experiments will further probe the early Universe processes that can generate GWs. 

The LIGO-Virgo detectors have observed dozens of GW signals from compact object mergers~\cite{LIGOScientific:2018mvr,LIGOScientific:2020ibl,LIGOScientific:2021djp}. From these observations we now have a good understanding of the population of $\mathcal{O}(10M_\odot)$ black holes (BHs) and their present merger rate~\cite{LIGOScientific:2020kqk,LIGOScientific:2021psn}. We can therefore estimate how strong astrophysical GW foreground these binary BHs (BBHs) generate. This foreground will impact the detectability of the primordial GW background with future GW experiments. We will present a simple analysis quantifying the effect of the BBH foreground on searches of the primordial GW background. 

Other astrophysical GW foregrounds arise from neutron star (NS) binaries, NS-BH binaries, and from binary white dwarfs (BWDs). The latter has been extensively studied in the literature~\cite{Nelemans:2001hp,Farmer:2003pa}, and we include its impact on the reach of GW experiments in our analysis as modeled for the LISA experiment in~\cite{Cornish:2017vip,Robson:2018ifk}. We neglect the contributions on the foreground arising from NS-NS and NS-BH binaries, as their population characteristics and merger rate is currently subject to large uncertainties and their contribution to the total GW foreground is subdominant at low frequencies to that arising from BBHs~\cite{LIGOScientific:2021psn}. However, they may still have an impact on the reach of the GW experiments as it is typically more difficult to resolve GW signals from NS-NS and NS-BH binaries than from BBHs.

The problem of probing the primordial GW background in the presence of an astrophysical foreground has been recently studied in several papers (see e.g.~\cite{Cornish:2001qi,Cutler:2005qq,Cornish:2007if,Harms:2008xv,Adams:2013qma,Regimbau:2016ike,Chen:2018rzo,Bartolo:2018qqn,Chatziioannou:2019zvs,Biscoveanu:2020gds,Sharma:2020btq,Sachdev:2020bkk,Boileau:2020rpg,Pieroni:2020rob,Flauger:2020qyi,Martinovic:2020hru,Littenberg:2020bxy,Cornish:2021smq,Boileau:2021sni,Boileau:2021gbr}). Our analysis complements the earlier studies by subtracting the resolvable BBHs from the total foreground and systematically analysing the detectability of a primordial GW background.\footnote{Subtracting the resolvable binaries from the foreground has been studied in the earlier literature~\cite{Cutler:2005qq,Regimbau:2016ike,Sharma:2020btq,Sachdev:2020bkk}, however  without systematical analysis of the detectability of the primordial GW background. Ref.~\cite{Biscoveanu:2020gds} presented a Bayesian implementation that estimates the astrophysical foreground and the detectability of the primordial GW background, but they demonstrate their method only in one benchmark case.} We account for the uncertainties in the BBH merger rate and perform a Fisher analysis to study the impact the BBH foreground on detection of primordial GW backgrounds modeled in the simplest case as power-laws. Moreover, we develop a new graphical presentation of the sensitivity of a given detector that accounts for the foreground. We find that the sensitivity of all experiments will be severely impacted by the BBH foreground, limiting their reach in probing primordial GW sources.

\vspace{4pt}\noindent\textbf{BBH merger rate} -- Assuming that the BH population that LIGO-Virgo is probing is astrophysical, we use a truncated power-law mass function and a merger rate that follows the star formation rate with a power-law delay time distribution distribution.\footnote{The modeling of the BBH population can be improved by considering different BBH formation channels~\cite{Zevin:2020gbd,Franciolini:2021tla}, but, in particular at low frequencies, the amplitude of the GW foreground from BBHs is well captured by the simple modeling that we use (cf.~\cite{Bavera:2021wmw}).} The differential BBH merger rate then is of the form
\bea \label{eq:RABH}
	&\frac{\td R(t)}{\td m_1 \td m_2} = \frac{R_0}{Z_\psi} M^\alpha \eta^\beta \psi(m_1)\psi(m_2) \\
	&\times \int \td t_d \td z_b P_b(z_b) P_d(t_d) \delta(t-t(z_b)-t_d) \,,
\eea
where the normalization factor $Z_\psi$ is defined such that $R = R_0$ at $z = 0$, $M=m_1+m_2$ and $\eta=m_1 m_2/M$ are the total mass and the symmetric mass ratio of the binary, and the mass dependence of the binary population is parametrized by $\alpha$ and $\beta$. The function $P_b(z)$ is given by the star formation rate~\cite{Madau:2016jbv},
\be
	{\rm SFR}(z) \propto \frac{(1 + z)^{2.6}}{1 + ((1 + z)/3.2)^{6.2}} \equiv P_b(z)\,,
\ee 
and for the time delay distribution we use $P_d(t) \propto t^{-1}$ at $t > 50\,\rm{Myr}$~\cite{Belczynski:2016obo}. The mass function is
\be
    \psi(m) \propto m^{\zeta} \, \theta(m - m_{\rm min})\theta(m_{\rm max}- m) \,,
\ee
with the normalization $\int \psi(m) \td\ln m = 1$. We take $m_{\rm min} = 3.0 M_\odot$ and $m_{\rm max} = 55 M_\odot$ corresponding to the estimates for the maximal mass of neutron stars and the beginning of the pair instability mass gap. We use the values $\alpha = 0$, $\beta = 6$, $\zeta = -1.5$ and $R_0 = 10_{-5}^{+6}\,{\rm Gpc}^{-3}{\rm yr}^{-1}$, following the fit performed in~\cite{Hutsi:2020sol} to the LIGO-Virgo observations.\footnote{Due to different modeling of the BBH population, this value of $R_0$ is lower than what is found e.g. in~\cite{LIGOScientific:2021psn}.} For indicative errors or the resulting stochastic GW foreground, we include the uncertainties only for the present merger rate $R_0$ and use the central values for the other parameters. This greatly simplifies our analysis.

\vspace{4pt}\noindent\textbf{Detectability of individual BBHs} -- The amplitude $|\tilde{h}(f)|$ of the Fourier transform of the inspiral-merger-ringdown GW signal from a BBH can be approximated as~\cite{Ajith:2007kx}\footnote{We use units where $c=1$ and we express all frequencies as measured today on the Earth.}
\bea
&|\tilde{h}(f)| = \sqrt{\frac{5\eta}{24}}\, \frac{\left[G M(1+z)\right]^{5/6}}{\pi^{2/3} D_L} \\&\times
\begin{cases}
f^{-7/6}\,, & f<f_{\rm merg}\,, \\
f_{\rm merg}^{-1/2} f^{-2/3}\,, & f_{\rm merg} \leq f < f_{\rm ring} \,,\\
f_{\rm merg}^{-1/2} f_{\rm ring}^{-2/3} \frac{\sigma^2}{4(f-f_{\rm ring})^2 + \sigma^2}\,, & f_{\rm ring} \leq f < f_{\rm cut}\,, \\
\end{cases}
\eea
where $D_L$ is the luminosity distance of the BBH, and the frequencies $f_{\rm merg}$, $f_{\rm ring}$, $f_{\rm cut}$ and $\sigma$ are of the form 
\be
    f_j = \frac{a_j \eta^2 + b_j \eta + c_j}{\pi GM(1+z)},
\ee 
with the coefficients $a_j$, $b_j$ and $c_j$ given in Table~I of Ref.~\cite{Ajith:2007kx}. 

The signal-to-noise ratio, ${\rm SNR}(f)$, of a BBH signal depends on the orientation of the detector with respect to the sky location of the binary and on the inclination of the binary. For an optimally oriented binary-detector system, neglecting the rotation of the detector, the signal-to-noise ratio of the BBH signal that in the end of the experiment is at frequency $f$, or the BBH has merged, is given by
\be \label{eq:SNR}
	{\rm SNR}(f) = \sqrt{4\int_{f(\mathcal{T} + \tau(f))}^f \td f'\, \frac{|\tilde{h}(f')|^2}{S_n(f')}} \,,
\ee
where $S_n$ is the noise power spectrum of the GW detector, that includes the instrumental noise and the GW foreground. The lower integration limit depends on the duration of the experiment, $\mathcal{T}$, and the coalescence time of the binary,
\be
 \tau(f) = \frac{5}{256} \frac{1}{\eta \,[(1+z) G M]^{5/3} (\pi f)^{8/3}}  \,.
\ee
We assume a duration of $\mathcal{T} = 4\,$yr for all experiments. 

We estimate how large fraction of the BBH signals can be resolved by calculating the probability~\cite{Finn:1992xs,Gerosa:2019dbe}
\be
p_{\rm det}(f) = \int_{r(f)}^1 p(\omega) \td \omega \,,
\ee 
where $r(f)={\rm SNR}_c/{\rm SNR}(f)$, that accounts for the antenna patterns $F_{+,\times}$ of the detector and averages over the binary sky location and inclination, and the polarization of the signal. The parameter $\omega$ is defined as
\be
\omega = \sqrt{\frac{(1+c_i^2)^2}{4} F_+(\theta,\phi,\psi)^2 + c_i^2 F_{\times}(\theta,\phi,\psi)^2} \,,
\ee
and its probability distribution $p(\omega)$ is calculated assuming uniform distributions for the binary inclination $c_i\in(-1,1)$, sky location $\cos\theta\in(-1,1)$ and $\phi\in(0,2\pi)$, and the polarization angle $\psi\in(0,2\pi)$. We use ${\rm SNR}_c = 8$ as the threshold signal-to-noise ratio.

\vspace{4pt}\noindent\textbf{GW foreground from LIGO-Virgo binaries} -- To calculate the GW foreground generated by the unresolvable BBH, we subtract from the total GW energy density emitted by all binaries the contribution from those that are sufficiently loud to be seen individually. The GW foreground then is different for different detectors, and its dimensionless energy density is (see e.g.~\cite{Zhu:2012xw,Phinney:2001di})
\be \label{eq:BBH}
	\Omega_{\rm BBH}(f) = \int\! \frac{\td V_c}{1+z} \td R(z) \frac{1}{\rho_c} \frac{\td \rho_{\rm GW}}{\td f} \left[1-p_{\rm det}(f)\right] ,
\ee
where $\td V_c$ is the differential comoving volume element at redshift $z$, $\rho_c$ is the critical energy density of the Universe, $\td \rho_{\rm GW}$ is the GW energy density emitted by a binary in the frequency range $(f,f+\td f)$~\cite{Phinney:2001di},\footnote{The factor $4/5$ accounts for the average over source orientations.}
\be
    \td \rho_{\rm GW} = \frac{4}{5}\frac{\pi}{G} \,f^3 |\tilde{h}(f)|^2 \td f \,,
\ee
$\td R$ is given by Eq.~\eqref{eq:RABH}, and the factor $1-p_{\rm det}(f)$ removes the binaries that are individually resolvable within the duration of the given experiment. Note that, in the analysis of actual data the estimated gravitational waveform for every detected event is subtracted from the time series data and the subtraction is affected by the errors in the parameter estimation. By comparing to the results obtained in~\cite{Sharma:2020btq} through the latter method, we find that our analysis is conservative regarding to how much the subtraction reduces the noise.

\begin{figure}
    \includegraphics[width=0.48\textwidth]{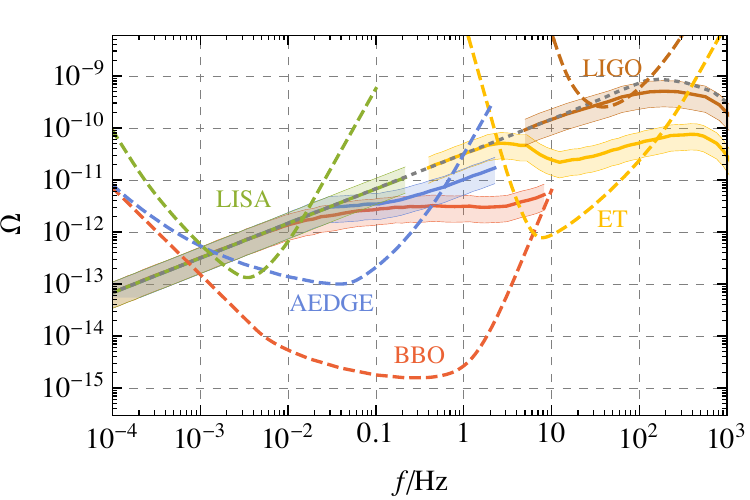} 
    \caption{
    Solid curves show the central value of the GW foreground from unresolvable BH binaries for different observatories, while the bands indicate their uncertainties stemming from uncertainty on BBH merger rate. Dashed curves show the reach of the corresponding observatories in terms of PI sensitivities, and the gray dotted curve indicates the total GW energy density emitted by the BH population. We assume $\mathcal{T} = 4\,$yr.}
    \label{fig:sgwb}
\end{figure}

In Fig.~\ref{fig:sgwb} we show the GW foreground from unresolvable BBHs for different GW experiments~\footnote{We limit the number of experiments shown for the sake of clarity, however, there are also other planned experiments including laser interferometers: Cosmic Explolor~\cite{Reitze:2019iox}, Taiji~\cite{Ruan:2018tsw} and TianQin~\cite{TianQin:2020hid}, B-DECIGO~\cite{Isoyama:2018rjb} and DECIGO~\cite{Yagi:2013du}, as well as experiments utilising atom interferometry: MAGIS~\cite{Abe_2021}, MIGA~\cite{Canuel_2018} ELGAR~\cite{Canuel:2019abg}, AION~\cite{Badurina:2019hst}, more akin to AEDGE~\cite{AEDGE:2019nxb,Canuel:2019abg,Badurina:2021rgt}.}.
The solid curves in Fig.~\ref{fig:sgwb} indicate the foreground for the central value of $R_0$. Since $\Omega_{\rm BBH} \propto R_0$ the boundaries of the $1\sigma$ band on the BBH foreground can be obtained by multiplying the central value result by $1.6$ and $0.5$ for the upper and lower limit. The gray dotted curve in Fig.~\ref{fig:sgwb} indicates the total GW foreground before any binaries are resolved, obtained by taking $p_{\rm det} \to 0$ in Eq.~\eqref{eq:BBH}. At $f\lesssim 100\,$Hz this curve is simply a power-law, $\Omega_{\rm BBH}\approx 3.2\times 10^{-11} (f/{\rm Hz})^{2/3}$. In Fig.~\ref{fig:sens} the BBH strain curve corresponds to the total BBH foreground for the central value of $R_0$. This is what we use as the BBH contribution to the instantaneous noise.

The dashed curves in Fig.~\ref{fig:sgwb} show the standard power-law integrated (PI) sensitivities~\cite{Thrane:2013oya} of the experiments, which account for the foregrounds only in the instantaneous detector noise. The signal-to-noise ratio of the GW background is given by~\cite{Allen:1997ad}
\be
    {\rm SNR}_{\rm BG} = \sqrt{ \mathcal{T} \int \td f \left[\frac{\Omega_{\rm tot}(f)}{\Omega_n(f)}\right]^2} \,,
\ee
and the PI curves are obtained as the envelope of the power-laws that give ${\rm SNR}_{\rm BG} = {\rm SNR}_c = 8$. The detector noise is the sum of the instrumental noise, and the GW foregrounds from BWDs (which we include as in Refs.~\cite{Cornish:2017vip,Robson:2018ifk}) and BBHs, $\Omega_n(f) = \Omega_{\rm instr}(f) + \Omega_{\rm BWD}(f) + \Omega_{\rm BBH}(f)$. Notice that here $\Omega_{\rm BBH}(f)$ is the total BBH foreground before any binaries are resolved.

We note that the GW foreground generated by BH binaries is not exactly smooth but includes fluctuations~\cite{Regimbau:2011rp,Renzini:2022alw}. At low frequencies, $f\lesssim 0.01$\,Hz, the foreground from solar mass BBHs consists of a large number of nearly-monochromatic signals and we expect the fluctuations to follow a narrow Gaussian distribution. At higher frequencies the fluctuations become larger and constitute so called 'popcorn' signal as the number of binaries contributing to the signal in a given frequency bin decreases. These fluctuations may help distinguishing this foreground from the primordial GW background. In this work we use the average GW foreground~\eqref{eq:BBH} and leave these improvements for future work.

\vspace{4pt}\noindent\textbf{Detectability of a primordial GW background} -- To study the impact of the BBH foreground on the detectability of the primordial GW background, we estimate how accurately the background can be measured using Fisher analysis~\cite{Fisher:1922saa} (see also e.g.~\cite{Cutler:1994ys,Balasubramanian:1995bm,Cutler:2005qq,Vallisneri:2007ev,Harms:2008xv,Sharma:2020btq,Sachdev:2020bkk,Pieroni:2020rob,Flauger:2020qyi,Gowling:2021gcy} for applications to GW data analysis). The total stochastic noise in the GW detector is given by the sum of the primordial GW background, which we model as a power-law, the WDB and the BBH foregrounds, and the instrumental noise,
\bea \label{eq:GWspectrum}
\Omega_{\rm tot}(f) =& \Omega \left(\frac{f}{f_{\rm ref}}\right)^{\alpha} + \,A\langle\Omega_{\rm BBH}(f)\rangle \\ &+ \Omega_{\rm BWD}(f) + \Omega_{\rm instr}(f)\,,
\eea
where $\Omega$ is the amplitude of the primordial GW background at a reference frequency $f_{\rm ref}$, $\alpha$ is the spectral index of the primordial GW background, and $A\langle\Omega_{\rm BBH}(f)\rangle$ is the contribution from unresolvable BBHs. We calculate $\langle\Omega_{\rm BBH}(f)\rangle$ using the central value for the present merger rate, $\langle R_0\rangle = 10\,{\rm Gpc}^{-3} {\rm yr}^{-1}$, and we have introduced the parameter $A=R_0/\langle R_0\rangle$ to account for the uncertainties in the amplitude of the BBH foreground. As our main aim is to demonstrate the impact of the BBH foreground on the detectability of a primordial GW background, for simplicity we don't include the uncertainties in the BWD foreground or in the instrumental noise in our analysis.\footnote{The uncertainties in the BWD foreground and in the instrumental noise, and their effects are considered previously in~\cite{Nelemans:2000es,Ruiter:2007xx,Nissanke:2012eh,Yu:2012tw,Littenberg:2014oda,Edwards:2015eka,Chatziioannou:2019zvs,Breivik:2019lmt,Edwards:2020tlp,Thiele:2021yyb} and \cite{Talbot:2020auc,Smith:2017vfk,Biscoveanu:2020kat,Littenberg:2013gja,Rover:2008yp,Rover:2011qd}.} Moreover,  simultaneously utilizing data from multiple experiments would in principle improve the prospects of detecting cosmological GW backgrounds~\cite{Barish:2020vmy}. However, in practice the relevant signals can be modeled as power-laws only locally~\cite{Caprini:2015zlo,Bartolo:2016ami,Caprini:2018mtu,Auclair:2019wcv,Caprini:2019egz} and in particular signals whose detectability in a given experiment depends on the impact of the foreground are not guaranteed to be visible in other experiments. For this reason we consider each experiment individually. 

The variances of the model parameters can be estimated from the inverse of the Fisher matrix, whose $ij$ component is given by
\be
\Gamma_{ij} = \mathcal{T} \int \td f\, \frac{\partial_i \Omega_{\rm tot}(f) \partial_j \Omega_{\rm tot}(f)}{\Omega_n(f)^2} \,,
\ee
as
\be
\sigma_j^2 = \Gamma^{-1}_{jj} \,.
\ee
Here the indices $i$ and $j$ label the parameters of the model, $i,j \in \{\Omega, \alpha, A\}$. Keeping $\alpha$ and $A$ fixed (that is, calculating the Fisher matrix only in $\Omega$) we get
\be
\frac{\sigma_\Omega}{\Omega} = \sqrt{\Omega^{-2} \, \Gamma_{\Omega\Omega}^{-1}} = {\rm SNR}_{\rm BG}^{-1} \,,
\ee
which shows that the amplitude of any power-law that touches the PI sensitivity curve can be measured with accuracy $\sigma_\Omega/\Omega = {\rm SNR}_c^{-1}$, assuming that the spectral index $\alpha$ and the BBH foreground are known with negligible uncertainties.

Next we aim to find a curve similar to the PI one, which takes into account the uncertainties in the BBH foreground. We perform a Fisher analysis in $\Omega$ and $A$ for different fixed values of $\alpha$ which we assume to know exactly as in the PI prescription. We account for the measured uncertainties in the present BBH merger rate $R_0$ by setting a prior on $A$ by the replacement
\be
\Gamma_{AA} \to \Gamma_{AA} + 1/\sigma_{A,0}^2 \,.
\ee
We use $\sigma_{A,0} = 0.6$, which arises from the $1\sigma$ upper bound $R_0 < 16\, {\rm Gpc}^{-3} {\rm yr}^{-1}$. 

\begin{figure}
    \includegraphics[width=0.48\textwidth]{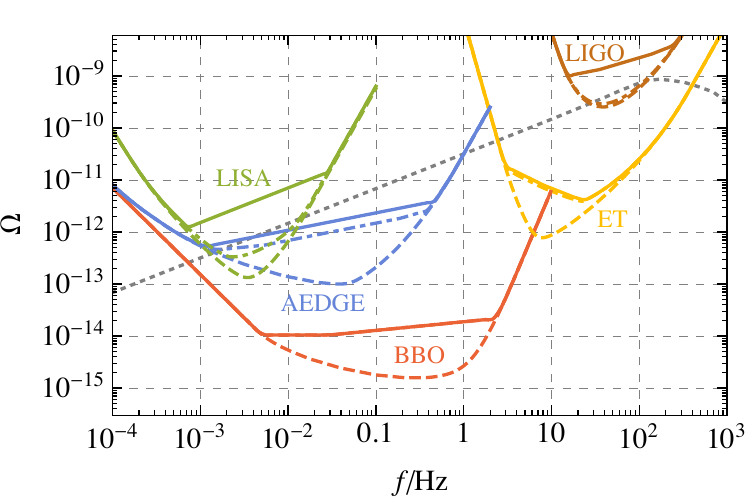} 
    \caption{Solid and dot-dashed curves show the sensitivities on primordial GW background accounting for the BBH foreground with the prior $\sigma_{A,0} = 0.6$ and $\sigma_{A,0} = 0.06$, respectively. Dashed curves show the PI sensitivities as in Fig.~\ref{fig:sgwb}.}
    \label{fig:omega}
\end{figure}

\begin{figure*}
    \includegraphics[width=\textwidth]{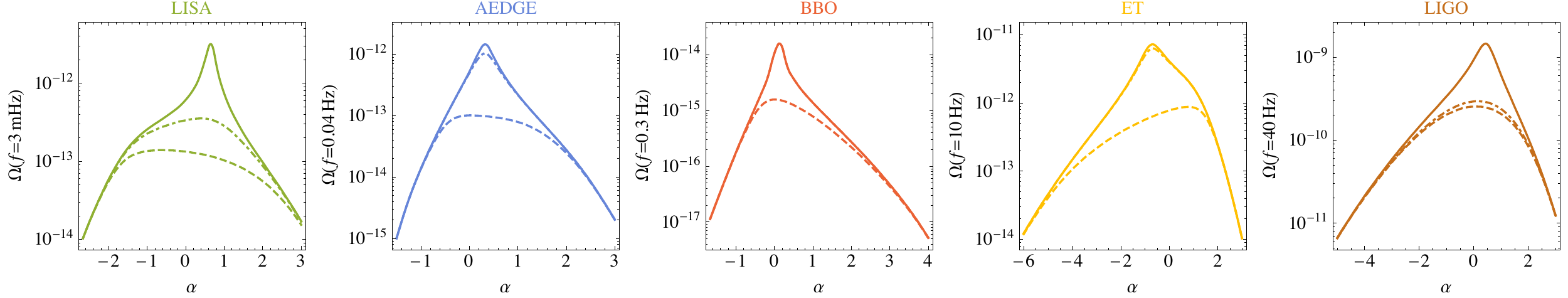} 
    \caption{Minimal detectable amplitude of the power-law primordial GW background as a function of its spectral index $\alpha$. The solid and dot-dashed curves include the BBH foreground with the current and an order of magnitude improved accuracy of the BBH merger rate, respectively. The dashed curves correspond to the usual PI sensitivities. We assume $\mathcal{T} = 4\,$yr.}
    \label{fig:alpha}
\end{figure*}

The solid curves in Fig.~\ref{fig:omega} show the sensitivities of different experiments obtained as described above by requiring that $\sigma_\Omega/\Omega < {\rm SNR}_c^{-1}$. In the same way as the PI curve, the curves shown in Fig.~\ref{fig:omega} depict the envelope of different power-laws that are detectable. As expected, these curves lie above the PI curves. The effect is particularly strong for LIGO, LISA and AEDGE, whereas ET and BBO can probe the primordial GW background well below the BBH foreground. The reason is that ET and BBO will be able to measure the BBH foreground very accurately even without an accurate prior knowledge of the BBH merger rate. Therefore also changing the prior of $A$ does not significantly change the sensitivity prospects for these experiments. Instead, for LISA and LIGO more accurate measurement of the BBH merger rate will greatly improve the prospects for probing primordial GW background. We show this by the dot-dashed curves that are calculated assuming an order of magnitude improvement in the accuracy of the BBH merger rate, which would imply $\sigma_{A,0} = 0.06$. For LIGO this curve overlaps with the PI sensitivity, whereas for BBO and ET it overlaps the curve obtained with the current accuracy on the rate, $\sigma_{A,0} = 0.6$.

\begin{table*}
\centering
\begin{tabular}{ p{8mm} p{8mm} | p{2cm} p{2cm} p{2cm} p{2cm} p{2cm} }
\hline\hline
$\sigma_{A,0}$ & $\sigma_{\alpha,0}$ & LISA & AEDGE & BBO & ET & LIGO \\
\hline
$0$ & $0$ & $1.3\times 10^{-13}$ & $9.9\times 10^{-14}$ & $1.6\times 10^{-15}$ & $7.6\times 10^{-13}$ & $2.5\times 10^{-10}$ \\
& $\infty$ & $1.4\times 10^{-13}$ & $1.0\times 10^{-13}$ & $1.6\times 10^{-15}$ & $1.0\times 10^{-12}$ & $2.5\times 10^{-10}$ \\
\hline
$0.06$ & $0$ & $3.3\times 10^{-13}$ & $4.6\times 10^{-13}$ & $1.0\times 10^{-14}$ & $3.9\times 10^{-12}$ & $2.9\times 10^{-10}$ \\
& $\infty$ & $3.5\times 10^{-13}$ & $1.3\times 10^{-12}$ & $2.3\times 10^{-14}$ & $5.1\times 10^{-12}$ & $3.0\times 10^{-10}$ \\
\hline
$0.6$ & $0$ & $5.9\times 10^{-13}$ & $4.8\times 10^{-13}$ & $1.0\times 10^{-14}$ & $4.0\times 10^{-12}$ & $9.3\times 10^{-10}$ \\
& $\infty$ & $2.0\times 10^{-12}$ & $3.3\times 10^{-12}$ & $2.3\times 10^{-14}$ & $5.4\times 10^{-12}$ & $1.5\times 10^{-9}$ \\
\hline\hline
\end{tabular}
\caption{Minimal detectable amplitude $\Omega$ of a flat ($\alpha=0$) primordial GW background. The first two columns show the priors used for the relative amplitude $A$ of the BBH foreground and the spectral index $\alpha$ of the primordial GW background.}
\label{table:flat}
\end{table*}

Fig.~\ref{fig:omega}, however, does not give a complete picture of the detectability of the primordial GW background. While any power-law that touches the sensitivity curves shown in Fig.~\ref{fig:omega} can be detected, for example with LISA assuming $\sigma_A = 0.6$ a flat primordial GW background can be detected even if it doesn't touch the solid green curve. As the foreground for LISA is a power-law with spectral index $2/3$, detecting a primordial GW background with $\alpha=2/3$ requires a larger amplitude, and therefore that power-law cuts the sensitivity curve for LISA in Fig.~\ref{fig:omega}. We show the minimal detectable amplitude for a given value of $\alpha$ at the peak sensitivity of each experiment in Fig.~\ref{fig:alpha}. For example, we see that for the primordial GW background with $\alpha = 2/3$ to be detectable by LISA, its amplitude needs to be higher than $3\times 10^{-12}$ at $3\,{\rm mHz}$, assuming the current uncertainties in the BBH merger rate.

Finally, for an easy comparison of different cases, in Table~\ref{table:flat} we show the values of minimal detectable amplitude of a flat, $\alpha=0$, primordial GW background, including also the case that the uncertainties in $\alpha$ are accounted for in the Fisher analysis. The latter case denoted with $\sigma_{\alpha,0} \to \infty$, and the case where $\alpha$ is fixed is denoted by $\sigma_{\alpha,0} \to 0$. When $\alpha$ is included in the Fisher analysis, the result depends also on the reference frequency $f_{\rm ref}$. In Table~\ref{table:flat} the reference frequencies for each experiment are fixed in the same way as in Fig.~\ref{fig:alpha}. In all cases the minimal detectable amplitude of the primordial GW background is determined by requiring that $\sigma_\Omega/\Omega < {\rm SNR}_c^{-1}$. For the first row in the table we assume that the foreground is known and calculate the Fisher matrix only in $\Omega$. As described above, this corresponds to the usual PI result.

\vspace{4pt}\noindent\textbf{Conclusions} -- We have analysed the impact of  the BH population currently probed by LIGO-Virgo on future searches for a primordial GW background. The GW foreground associated with the current observations has not been probed directly yet. However, the rate of events inferred by the LIGO-Virgo observations suggests that it will have a large amplitude, and will cut significantly into the integrated sensitivities of upcoming experiments, limiting their reach. 

We have estimated the BBH foreground by subtracting from the total GW energy density emitted by BBHs the contribution from the binaries that can be resolved individually. Then, using Fisher analysis, we have assessed the impact of this foreground on future searches of a primordial GW background. Describing the primordial background as a power-law with a given slope, we have checked what amplitudes could be probed in the presence of the foreground with each experiment. We have found the impact to be much smaller than one would expect from simply looking at the amplitude of the foreground, especially for detectors that can themselves probe the foreground accurately. Instead for detectors unable to significantly improve the foreground measurement the final result depends on priors. We have illustrated this by showing also results assuming an order of magnitude improvement in the accuracy of the BBH merger rate. In the case of LISA and LIGO this significantly alleviates the problem, while giving almost negligible change in AEDGE, BBO and ET. However, while subtraction of the loud identifiable binaries and improvements in the measurement of the BBH merger rate help, the sensitivity of all future experiments in terms of the abundance of the primordial background they can reach is lowered in each case.

\begin{acknowledgments}
\vspace{4pt}\noindent\emph{Acknowledgments} -- This work was supported by the Spanish MINECO grants FPA2017-88915-P and SEV-2016-0588, the Spanish MICINN grants IJC2019-041533-I and PID2020-115845GB-I00/AEI/10.13039/501100011033, the grant 2017-SGR-1069 from the Generalitat de Catalunya, the Polish National Science Center grant 2018/31/D/ST2/02048, and the Polish National Agency for Academic Exchange within Polish Returns Programme under agreement PPN/PPO/2020/1/00013/U/00001. IFAE is partially funded by the CERCA program of the Generalitat de Catalunya.
\end{acknowledgments}

\bibliography{refs}

\end{document}